\documentclass[prb,twocolumn,showpacs,preprintnumbers,amsmath,amssymb]{revtex4}

\usepackage{graphicx}
\usepackage{dcolumn}
\usepackage{bm}

\begin{document}

\title{Quantum Hall resistances of multiterminal top-gated graphene device}

\author{Dong-Keun Ki}
 \affiliation{Department of Physics, Pohang University of Science
and Technology, Pohang 790-784, Korea}

\author{Hu-Jong Lee}
\email{hjlee@postech.ac.kr} \affiliation{Department of Physics,
Pohang University of Science and Technology, Pohang 790-784, Korea}
\affiliation{National Center for Nanomaterials Technology, Pohang
790-784, Republic of Korea}

\date{\today}

\begin{abstract}
Four-terminal resistances, both longitudinal and diagonal, of a
locally gated graphene device are measured in the quantum-Hall
(QH) regime. In sharp distinction from previous two-terminal
studies [J. R. Williams \textit{et al.}, Science {\bf 317}, 638
(2007); B. \"{O}zyilmaz \textit{et al.}, Phys. Rev. Lett. {\bf
99}, 166804 (2007)], asymmetric QH resistances are observed, which
provide information on reflection as well as transmission of the
QH edge states. Most quantized values of resistances are well
analyzed by the assumption that all edge states are equally
populated. Contrary to the expectation, however, a 5/2
transmission of the edge states is also found, which may be caused
by incomplete mode mixing and/or by the presence of
counter-propagating edge states. This four-terminal scheme can be
conveniently used to study the edge-state equilibration in locally
gated graphene devices as well as mono- and multi-layer graphene
hybrid structures.
\end{abstract}

\pacs{73.43.Fj, 71.70.Di, 73.61.Wp, 73.23.-b}

\maketitle

\section{INTRODUCTION}

Landau-level splitting in two-dimensional (2D) electron systems
under a perpendicular magnetic field reveals the well-known
quantum-Hall (QH) effect.~\cite{Klitzing80,Beenakker91} When the
Fermi energy is set between two Landau levels, a current circulates
along the edge conduction states in a (chiral) direction determined
by the carrier type and the direction of the magnetic
field.~\cite{Halperin82,Buttiker88,Beenakker91} Utilizing this
chiral character of the edge states one can devise diverse
solid-state beam splitters out of 2D electron gas
systems.~\cite{Oliver99,Henny99,Ji03}

On the other hand, due to the relativistic nature of the carriers,
graphene, a 2D honeycomb lattice of carbon atoms, shows the
half-integer QH
effect.~\cite{Novoselov04,Novoselov05,Zhang05,Neto09} Thus, the
manipulation of the edge states in graphene can be of particular
interest. However, the electrostatic deflection of the edge
state~\cite{Beenakker91,Oliver99,Henny99,Ji03} is not realizable
in graphene due to the Klein tunneling of the carriers through an
electrostatic barrier.~\cite{Katsnelson06} Nonetheless, the
possibility of controlling the edge-state transmission in graphene
has been confirmed by locally modulating the filling factor $\nu$
and the chiral direction of the edge
states.~\cite{Ozyilmaz07,Williams07} Two-terminal observation of
the edge-state transmission in graphene to date is well explained
by the complete-mode-mixing hypothesis where all edge states are
equally populated at \textit{p-n}
interfaces.~\cite{Ozyilmaz07,Williams07,Abanin07} QH plateaus in
two-terminal measurements, however, can be distorted depending on
the sample geometry and the contact
inhomogeneities.~\cite{Abanin08,Williams08} Moreover, the
two-terminal conductance gives the information only on the
edge-state transmission, lacking the information on the
reflection. Thus, for the better manipulation of the edge states
of an arbitrary-shaped graphene device, one needs
geometry-independent measurements that can furnish information on
both the edge-state reflection and transmission.

\begin{figure}[b]
\begin{center}
\leavevmode
\includegraphics[width=0.7\linewidth]{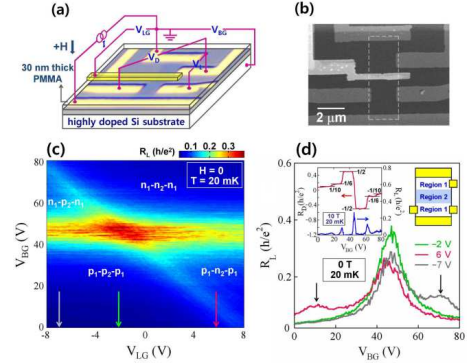}
\caption{(Color) (a) Schematic measurement configuration. (b)
Scanning electron microscope image of the device after depositing
a local gate, the brightest central part. The broken lines
represent the graphene edges and the boundary of the PMMA
insulation layer is evident by the contrast change near the top of
the image. The white scale bar represents 2 $\mu$m. (c)
Two-dimensional color map of $R_L$($V_{LG}$,$V_{BG}$) for 0 T at
20 mK. A set of carrier types in the regions 1 and 2 is labeled in
each quadrant. (d) $V_{BG}$ dependence of $R_L$ at $V_{LG}$=-2 V,
6 V and -7 V extracted from (c) as indicated by arrows with the
same color (green, red, and gray, respectively). Left inset: the
$R_D$ (upper) and $R_L$ (lower) as a function of $V_{BG}$ at
$V_{LG}$= -2 V and 10 T, showing the half-integer QH effect. Right
inset: a schematic top view of the device, where the local gate is
placed on top of the colored region at the center (the region 2).}
\end{center}
\end{figure}

In this paper, we report on four-terminal QH transport
measurements in a top-gated bi-polar graphene device, which show
the quantization of longitudinal QH resistances as well as an
asymmetry in the diagonal QH resistances (the meaning will be
defined below). Our measurement scheme provides precise
information on the reflected QH edge states in addition to the
transmitted ones as can be obtained from two-terminal
measurements.~\cite{Ozyilmaz07,Williams07} Most of the results are
in good quantitative agreement with the Landauer-B\"{u}ttiker
formula,~\cite{Beenakker91,Buttiker88} while unexpected resistance
plateaus corresponding to the 5/2 transmission of edge state are
also observed. It may arise from the incomplete mode mixing and/or
unusual QH edge states that are possibly present under the local
gate.~\cite{Chklovskii92,Rossier07,Silvestrov08,Zhang06,Abanin07-1,Jiang07,Abanin07-2}
This simple four-terminal scheme will allow an additional insight
into the edge-state equilibration in bipolar graphene systems.
Results of this study can also be utilized for the spatial
manipulation of the Dirac fermions in graphene, which is one of
the hot issues in graphene studies.

\begin{figure}[t]
\begin{center}
\leavevmode
\includegraphics[width=0.7\linewidth]{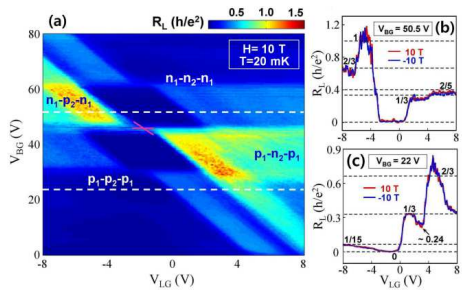}
\caption{(Color) (a) Two-dimensional plot of
$R_L$($V_{LG}$,$V_{BG}$) for 10 T at 20 mK. A set of carrier types
in the regions 1 and 2 is labeled in each quadrant. (b,c) Plots of
$R_L$($V_{LG}$) extracted from (a) at $V_{BG}$=50.5 V and 22 V, as
indicated by the white broken lines in (a). Broken lines represent
the calculation result based on the Landauer-B\"{u}ttiker
formula.}
\end{center}
\end{figure}

\section{SAMPLE PREPARATION and VERIFICATION}

The sample was prepared by mechanically exfoliating monolayer
graphene~\cite{Novoselov04} on a silicon substrate covered with a
300-nm-thick oxide layer where the silicon substrate was used as a
back gate (BG). Electrical contacts of Cr (5 nm)/Au (20 nm) were
patterned by the method described elsewhere.~\cite{Ki08} It was
then followed by spin-coating a 30-nm-thick polymethyl
methacrylate (PMMA, 950 K, 2{\%} in anisole) dielectric layer on
top of the device, which was cross-linked by high doses (15000
$\mu$C/cm$^2$) of electron beams~\cite{Huard07} with 20 keV.
Finally, a local gate (LG) of Cr (5 nm)/Au (40 nm) was deposited
at the center of the device [Fig. 1(b)]. The device was cooled
down to 20 mK in a dilution fridge (Oxford Instruments, Model
AST). The longitudinal ($R_L$=$V_L$/$I$) and diagonal
($R_D$=$V_D$/$I$) resistances were measured simultaneously with
two lock-in amplifiers, synchronized with each other at $I$=2 nA
and $f$=13.3 Hz [see Fig. 1(a)].

Figure 1(c) shows a 2D color map of $R_L$($V_{LG}$,$V_{BG}$)
measured at zero magnetic field in units of $h$/$e^2$ (all
resistances will be presented in this unit afterward). In the
figure, horizontal and diagonal bands crossing at
($V_{LG}$,$V_{BG}$)$\approx$(-2 V,46 V) are identified, which
divide the map into four different quadrants. These bands
represent positions of local resistance maxima arising from the
change in the carriers in the region underneath the local gate
(the region 2) with respect to the outside region (the region 1)
[the right inset of Fig. 1(d)]. Figure 1(d) is the one-dimensional
slice plot of $R_L$($V_{BG}$) that is extracted from Fig. 1(c) at
$V_{LG}$=-2 V, 6 V, and -7 V. It indicates that the position of
the secondary resistance peak, corresponding to the diagonal
pattern in Fig. 1(c), directly depends on $V_{LG}$ while that of
the dominant resistance peak, the horizontal pattern in Fig. 1(c),
is almost insensitive to $V_{LG}$. This leads to a conclusion that
the horizontal (diagonal) band is from the charge-neutrality point
in the region 1 (in the region 2) where the carrier type is
altered. It verifies the successful performance of our bipolar
device.~\cite{Ozyilmaz07,Williams07,Huard07,Gorbachev08,Stander09}
Additionally, in the left inset of Fig. 1(d), we show the
half-integer QH effect, $R_D$ (upper) and $R_L$ (lower), taken at
$V_{LG}$=-2 V and $B$=10 T, indicating that our device consists of
a monolayer graphene sheet.~\cite{Novoselov05,Zhang05,Neto09}

\begin{figure}[t]
\begin{center}
\leavevmode
\includegraphics[width=0.7\linewidth]{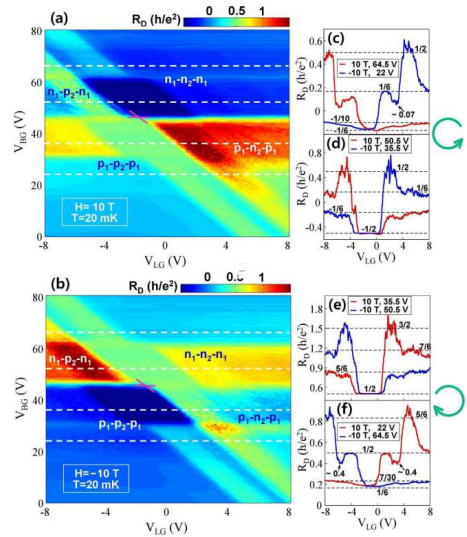}
\caption{(Color) [(a) and (b)] Two-dimensional plots of
$R_D$($V_{LG}$,$V_{BG}$) at 10 T and -10 T, respectively. A set of
carrier types in the regions 1 and 2 is labeled in each quadrant.
[(c)-(f)] Red curves: $R_D$($V_{LG}$) extracted from (a) at
$V_{BG}$=64.5 V, 50.5 V, 35.5 V, and 22 V. Blue curves:
$R_D$($V_{LG}$) extracted from (b) at $V_{BG}$=22 V, 35.5 V, 50.5
V, and 64.5 V. Thus, (c) and (d) correspond to the
\textit{counterclockwise} edge state in the region 1, while (e)
and (f) to the \textit{clockwise} edge state in the same region as
denoted on the right. The extraction $V_{BG}$ values are marked by
the white broken lines in (a) and (b). Broken lines represent the
calculation results from the Landauer-B\"{u}ttiker formula.}
\end{center}
\end{figure}

\section{EXPERIMENTAL RESULTS}

The 2D color map of $R_L$($V_{LG}$,$V_{BG}$) measured at 10 T is
shown in Fig. 2(a). It displays several skewed blocks of different
resistances, implying the quantization of the longitudinal
resistance. Details are more clearly seen in one-dimensional
slices, $R_L$($V_{LG}$), of Fig. 2(a) for fixed values of
$V_{BG}$. In Figs. 2(b) and 2(c), $R_L$($V_{LG}$) is displayed for
$V_{BG}$=50.5 V and 22 V, respectively. The red and blue curves
were taken at 10 T and -10 T, respectively. Two curves almost
completely overlap with each other. Dominant resistance plateaus
exist at zero resistance in the region -3 V$<$$V_{LG}$$<$1 V [Fig.
2(b)] and -3 V$<$$V_{LG}$$<$0 V [Fig. 2(c)], which arise from the
full transmission of the edge states when the filling factors in
the regions 1 ($\nu_1$) and 2 ($\nu_2$) are identical. In addition
to these trivial ones, one also finds plateaus of non-zero
fractional resistances such as 1/15, 1/3, and 2/3 in a certain
range of $V_{LG}$. This quantization directly demonstrates that a
portion of the edge states is reflected at the interfaces between
the regions 1 and 2 for non-identical filling factors $\nu_1$ and
$\nu_2$, which is consistent with the previous two-terminal
conductance measurements.~\cite{Ozyilmaz07,Williams07} The
quantized values of $R_L$ are in excellent agreement with the
calculation results following Ref. [4] as represented by the
broken lines in Figs. 2(b) and 2(c), the details of which will be
discussed below.

Now, let us focus on the diagonal resistance $R_D$, which exhibits
far richer features. As seen in Figs. 3(a) and 3(b), the 2D plots
of $R_D$($V_{LG}$,$V_{BG}$) taken at 10 T and -10 T, respectively,
also reveal skewed blocks of different resistances, but with
overall features much different from Fig. 2(a). First of all, both
Figs. 3(a) and 3(b) show no inversion symmetry with respect to the
crossing point at $V_{BG}$$\sim$46 V and $V_{LG}$$\sim$-2 V, where
$\nu_1$=$\nu_2$=0 (the zero point). Nonetheless, there exists an
inversion symmetry between Figs. 3(a) and 3(b), or equivalently a
180$^{\circ}$ rotational symmetry between them with respect to the
zero point, which is in sharp distinction from the feature of
$R_L$ [Fig. 2(a)] as well as the previous two-terminal
results.~\cite{Ozyilmaz07,Williams07} More details are revealed by
the one-dimensional slices shown in Figs. 3(c)-3(f), which are
again extracted from Fig. 3(a) for 10 T (red curves) and Fig. 3(b)
for -10 T (blue curves). The corresponding values of $V_{BG}$ are
specified in each figure. The figures illustrate that the data
taken at 10 T and -10 T are mirror symmetric with respect to
$V_{LG}$$\sim$-1 V, which again confirms the inversion symmetry of
$R_D$ between the two opposite field directions.

Figures 3(e) and 3(f) show the $R_D$ measured when carriers in the
region 1 are holes ($\nu_1$$<$0) for 10 T and electrons
($\nu_1$$>$0) for -10 T, which corresponds to the
\textit{clockwise} edge-states in the region 1. Most of the
quantized resistances in these figures match with the inverse of
two-terminal conductance observed previously.~\cite{Ozyilmaz07}
Thus, $R_D$ in this region corresponds to the two-terminal (or
Hall) resistance of transmitted edge states. In contrast to the
two-terminal results, however, Figs. 3(e) and (f) show clear 1/2
plateaus. This indicates that the disorder effect, which has been
regarded as the cause of the observed reduction in the conductance
for $\nu_1$=$\nu_2$=$\pm$2 in the previous
study,~\cite{Ozyilmaz07} is negligible in this four-terminal
measurement. On the other hand, in Figs. 3(c) and 3(d) (for
$\nu_1$$>$0 at 10 T and $\nu_1$$<$0 at -10 T), $\pm$1/2, $\pm$1/6,
and -1/10 plateaus are seen together with a sign change at certain
values of $V_{LG}$. This sign change seems to be odd because these
plateaus correspond to the \textit{counterclockwise} edge states
in the region 1, where $R_D$ is supposed to be negative. A careful
analysis, however, indicates that the $R_D$ for $\nu_1$$\cdot$$\nu
_2$$<$0 [the n$_1$-p$_2$-n$_1$ region in Fig. 3(a) and the
p$_1$-n$_2$-p$_1$ region in Fig. 3(b)] is nothing but the Hall
resistance for the \textit{clockwise} edge states in the region 2,
which should be positive as observed. But, still counterintuitive
positive (1/6) plateaus appear in Fig. 3(c) even for the
\textit{counterclockwise} edge states both in the regions 1 and 2
[$\nu_1$$\cdot$$\nu_2$$>$0 for the n$_1$-n$_2$-n$_1$ region in
Fig. 3(a) and for the p$_1$-p$_2$-p$_1$ region in Fig. 3(b)]. This
can be accounted for in terms of the Landauer-B\"{u}ttiker
formalism for the edge states as will be shown
below.~\cite{Beenakker91,Buttiker88}

\section{DISCUSSION}

\begin{figure}[t]
\begin{center}
\leavevmode
\includegraphics[width=0.7\linewidth]{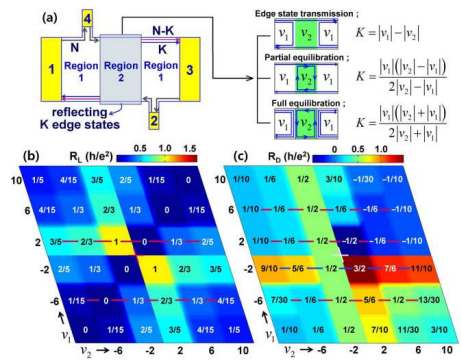}
\caption{(Color) (a) Schematic top view of the edge-state
configuration. Electrical contacts are shown in yellow and thick
arrows represent the chiral direction of QH edge states. The case
of the \textit{clockwise} circulation in region 1 is shown as an
example. We assume that $K$ out of $N$ edge states are reflected
at the interfaces between the regions 1 and 2 (blue arrows). On
the right, three possible equilibration processes in a graphene
\textit{p-n-p} device and corresponding values of $K$ for each
regime are illustrated. [(b) and (c)] Two-dimensional plot of the
calculated $R_L$ and $R_D$ for a positive magnetic field as a
function of $\nu_1$ and $\nu_2$. In both figures, calculated
values of the resistance plateaus are written at the center of
each block in units of $h$/$e^2$. Horizontal lines represent the
filling factor $\nu_1$ where one-dimensional data in Figs. 2 and 3
are extracted.}
\end{center}
\end{figure}

The Landauer-B\"{u}ttiker formula~\cite{Beenakker91,Buttiker88} is
suitable for studying the one-dimensional edge-state transport
where the current through the lead $\alpha$ ($I_{\alpha}$) is
expressed as a linear combination of transmission coefficients
($T_{\alpha\beta}$) multiplied by the corresponding chemical
potentials ($\mu_{\beta}$). Accordingly, it is required to
evaluate the scattering matrix ($\widehat{S}$) with the elements
$T_{\alpha\beta}$'s and then solve the linear equations
($\overrightarrow{I}$=$\widehat{S}$$\overrightarrow{\mu}$), which
is straightforward in our case because the reflection ($K$)
happens at \textit{p-n} interfaces only [see Fig. 4(a)]. This
procedure is adopted to obtain $R_L$ and $R_D$ in Ref. [4] for the
configuration that is identical to ours. It is suggested that
$R_L$ is finite unless $K$ is zero and $R_D$ has two different
values ($R_D^+$ and $R_D^-$), depending on the measurement
configurations.

\begin{eqnarray}
\label{eq1}
 R_{1342}&\rightarrow& \frac{h}{e^2}\frac{1}{N-K}\equiv R_{D}^{+}, \nonumber \\
 R_{4213}&\rightarrow& -\frac{h}{e^2}\frac{N-2K}{N(N-K)}=R_{D}^{+}-\frac{h}{e^2}\frac{2}{N}\equiv R_{D}^{-}, \\
 R_{1243}&=&R_{4312}\rightarrow \frac{h}{e^2}\frac{K}{N(N-K)}=R_{D}^{+}-\frac{h}{e^2}\frac{1}{N}\equiv
 R_L, \nonumber
\end{eqnarray}
\\
\noindent where $R_{\alpha \beta \gamma \delta}$ represents the
resistance measured between the voltage leads $\gamma$ and
$\delta$ for the current injection from leads $\alpha$ to $\beta$.
In Eq. (\ref{eq1}) it is assumed that $N$ edge states circulate in
the \textit{clockwise} direction in the region 1, i.e.,
$\nu_1$$<$0 ($>$0) for 10 T (-10 T) in our case. For a direct
comparison with the theoretical expectation,~\cite{Buttiker88} we
adopt a four-terminal schematic sample configuration in Fig. 4(a)
rather than a five-terminal one corresponding to the actual
device. Since each resistance ($R_L$ or $R_D$) was obtained in a
four-terminal configuration, the schematic is equivalent to the
real device configuration, namely, $R_D$ ($R_L$) in Fig. 1(a)
corresponds to $R_{1342}$ ($R_{4312}$) in Fig. 4(a). Since
$R_{1342}$ and $R_{4213}$ depend on the chiral direction of
incoming edge states $R_D$ corresponds to $R_{D}^{+}$
($R_{D}^{-}$) for the \textit{clockwise}
(\textit{counterclockwise}) edge states in the region 1. Thus,
$R_D$ changes as the sign of $\nu_1$ changes at a fixed
magnetic-field direction, but it represents the same resistance if
both $\nu_1$ and the magnetic-field direction are altered. This
explains the inversion symmetry shown in Fig. 3.

Two-terminal resistance in a bipolar configuration studied
previously~\cite{Ozyilmaz07} corresponds to $R_{D}^{+}$ only in
Eq. (\ref{eq1}), which is the Hall resistance from the $N-K$
transmitted edge states. It is consistent with our observation
shown in Figs. 3(e) and 3(f), which correspond to the
\textit{clockwise} edge states in the region 1. But, the
information on the reflected edge states is lost in the
two-terminal process. On the other hand, $R_{D}^{-}$ represents
the eliminated Hall resistance by the $K$ reflected edge-states,
which can be observed only in a four-terminal configuration. This
implies that the average of $R_{D}^{+}$ and $R_{D}^{-}$,
($R_{D}^{+}$$-$$R_{D}^{-}$)/2, provides the Hall resistance of $N$
incoming edge states (the sign for $R_{D}^{-}$ is reversed because
$R_{D}^{+}$ and $R_{D}^{-}$ are measured in the opposite direction
with respect to the current flow). Thus, if more than a half of
the incoming edge states are reflected (2$K$$>$$N$), $R_{D}^{-}$
can change the sign, accounting for the odd sign of resistances
seen in Figs. 3(c) and 3(d). On the other hand, $R_L$ indicates
the difference between the (two-terminal) Hall resistance of the
$N-K$ transmitted and the $N$ incoming edge states which is always
positive or equal to zero for $K$=0. In consequence, all the
unexpected features in $R_D$ as well as the finite value of $R_L$
are well accounted for at least qualitatively.

For a quantitative analysis, we estimated the number of the
reflected edge states ($K$) in three different
regimes:~\cite{Ozyilmaz07} the edge-state-transmission regime
($\nu_1$$\cdot$$\nu_2$$>$0, $N$=$|\nu_1|$$\geq$$|\nu_2|$),
partial-equilibration regime ($\nu_1$$\cdot$$\nu_2$$>$0,
$N$=$|\nu_1|$$<$$|\nu_2|$), and full-equilibration regime
($\nu_1$$\cdot$$\nu_2$$<$0, $N$=$|\nu_1|$). Based on the
complete-mode-mixing hypothesis,~\cite{Ozyilmaz07,Abanin07} it is
evident that $K$ is equal to $|\nu_1|$-$|\nu_2|$ in the
edge-state-transmission regime, where the extra edge states in the
region 1 ($|\nu_1|$-$|\nu_2|$) are forbidden in the region 2. In
the partial-equilibration regime, the excess states
($|\nu_2|$-$|\nu_1|$) circulate in the region 2 while in partial
equilibration with the incoming edge states ($|\nu_1|$). Finally,
in the full-equilibration regime, the edge states in the regions 1
($|\nu_1|$) and 2 ($|\nu_2|$) circulate in opposite directions,
propagating in the same direction along the \textit{p-n}
interfaces and equilibrate with each other. In the last two
regimes, one can calculate $K$ by using the current-conservation
relation.~\cite{Ozyilmaz07} Schematic edge-state circulation and
values of the $K$ are illustrated in the right panel of Fig. 4(a)
for each regime. With these values of filling factor in Eq.
(\ref{eq1}), the $R_L$ and $R_D$ are calculated as functions of
$\nu_1$ and $\nu_2$ as shown below.
\\
\\
\noindent (a) Edge-state transmission regime:
\begin{eqnarray}
R_D &=& \begin{cases} \frac{h}{e^2}\frac{1}{|\nu_2|} &
\text{($\nu_1$$<$0 at 10 T or $\nu_1$$>$0 at -10 T)} \\
-\frac{h}{e^2}\frac{2|\nu_2|-|\nu_1|}{|\nu_1||\nu_2|} &
\text{($\nu_1$$>$0 at 10 T or $\nu_1$$<$0 at -10 T)}
\end{cases}, \nonumber \\
R_L &=& \frac{h}{e^2}\frac{|\nu_1|-|\nu_2|}{|\nu_1||\nu_2|}
\end{eqnarray}
\\
\\
\noindent (b) Partial equilibration regime:
\begin{eqnarray}
R_D &=& \begin{cases}
\frac{h}{e^2}\frac{2|\nu_2|-|\nu_1|}{|\nu_1||\nu_2|} &
\text{($\nu_1$$<$0 at 10 T or $\nu_1$$>$0 at -10 T)} \\
-\frac{h}{e^2}\frac{1}{|\nu_2|} & \text{($\nu_1$$>$0 at 10 T or
$\nu_1$$<$0 at -10 T)}
\end{cases}, \nonumber \\
R_L &=& \frac{h}{e^2}\frac{|\nu_2|-|\nu_1|}{|\nu_1||\nu_2|}
\end{eqnarray}
\\
\\
\noindent (c) Full equilibration regime:
\begin{eqnarray}
R_D &=& \begin{cases}
\frac{h}{e^2}\frac{2|\nu_2|+|\nu_1|}{|\nu_1||\nu_2|} &
\text{($\nu_1$$<$0 at 10 T or $\nu_1$$>$0 at -10 T)} \\
\frac{h}{e^2}\frac{1}{|\nu_2|} & \text{($\nu_1$$>$0 at 10 T or
$\nu_1$$<$0 at -10 T)}
\end{cases}, \nonumber \\
R_L &=& \frac{h}{e^2}\frac{|\nu_1|+|\nu_2|}{|\nu_1||\nu_2|}
\end{eqnarray}
\\
\noindent Results for a positive magnetic field (10 T) are
summarized in two color maps in Figs. 4(b) and 4(c), respectively,
with the calculated resistance labeled in each block. As seen in
Fig. 4(b), unless $\nu_1$=$\nu_2$, $R_L$ becomes positive and is
in good quantitative agreement with the observed results as
denoted by the broken lines in Figs. 2(b) and 2(c). The values of
$R_D$ are also well reproduced by the calculation as shown in Fig.
4(c) as well as Fig. 3(a). Although not shown, the observed -10 T
data of $R_D$ [Fig. 3(b)] are also well fit by the calculation.
Furthermore, for $\nu_1$$>$0 ($<$0) at 10 T (-10 T), the
calculated $R_D$ becomes positive in the region where
$\nu_1$$\cdot$$\nu_2$$<$0 or $|\nu_1|$$>$2$|\nu_2|$ and
$\nu_1$$\cdot$$\nu_2$$>$0, reproducing the odd sign of resistances
found in Figs. 3(c) and 3(d). Again, all the observed features of
$R_D$ are well accounted for in terms of the Landauer-B\"{u}ttiker
formula.~\cite{Beenakker91,Buttiker88} Since all of these results
are linked to the chiral-directional dependence of $R_D$, one can
conclude that edge states in graphene also behave in the same way
as those in a conventional 2D electron gas,~\cite{Beenakker91}
although the QH effect itself is distinct by the Dirac nature of
charge carriers.~\cite{Novoselov05,Zhang05,Neto09}

\section{5/2 TRANSMISSION of EDGE STATES}

Most of the observed results in the color maps shown in Figs. 2 and
3 are quantitatively accounted for. However, one can distinguish
some additional features which appear as a faint diagonal bands
crossing from the bottom-right region to the top-left region of the
figures in the ranges $V_{BG}$$>$$\sim$60 V and $V_{BG}$$<$$\sim$30
V near the boundary where $\nu_2$ changes sign. In fact, these band
structures constitute additional resistance plateaus of $\sim$0.24,
$\sim$0.07, and $\sim$0.4 as indicated by arrows in Figs. 2(c),
3(c), and 3(f), respectively. Solving Eq. (\ref{eq1}) for
$R_L$$\sim$0.24, $R_{D}^{-}$$\sim$0.07, and $R_{D}^{+}$$\sim$0.4 at
$N$=$|\nu_1|$=6 gives almost identical $K$ values as $\sim$3.54,
$\sim$3.52, and $\sim$3.5, respectively. This indicates that these
unexpected resistance plateaus are all produced by a single 5/2
(=$N$$-$$K$=6$-$7/2) edge-state transmission for $|\nu_1|$=6,
$|\nu_2|$=2, and $\nu_1$$\cdot$$\nu_2$$>$0, just before $\nu_2$
changes sign or, equivalently, just before the edge-state
transmission regime turns into the full equilibration regime. This
state may be related to the disorder in the region 2, but no related
features are present at $\nu_1$=$\nu_2$=$\pm$2 in our
measurement.~\cite{Ozyilmaz07} In addition, the disorder effect is
not supposed to be significant in our four-terminal measurement.

Another possible explanation is based on the fact that the plateaus
take place near the boundary between the edge-state transmission and
the full-equilibration regimes. One may imagine the presence of
counter-propagating edge states: for instance, $|2|$+$|\varepsilon|$
\textit{clockwise} edge states and $|\varepsilon|$
\textit{counterclockwise} edge states in the region 2. Here, we
expect that $|\varepsilon|$ should be an integer less than 2,
because no fractional QH effect is present~\cite{Hou07,Chamon08} and
the maximum filling factor in the region 2 is two. However, a simple
calculation of the total reflection $K$ by adding two different $K$
values for the \textit{clockwise} edge states (edge-state
transmission) and the \textit{counterclockwise} ones (full
equilibration) leads to $|\varepsilon|$ much larger than 2 (about
4.7, corresponding to $K$=7/2), which is far from the expectation.

The discrepancy can arise from the incomplete mode mixing for the
counter-propagating edge states, which leads to the 5/2 transmission
when only two and a half out of three \textit{clockwise} edge states
participate in the equilibration process completely. These
counter-propagating edge states may originate from the compressible
and incompressible edge
states~\cite{Chklovskii92,Rossier07,Silvestrov08} formed near the
boundary of the graphene by the charge accumulation as suggested for
the graphene sheet placed above a global
backgate.~\cite{Rossier07,Silvestrov08} However, the strange
diagonal bands observed in our study indicate that the local gate,
\emph{by some means, further enhances the charge accumulation.} The
presence of the QH ferromagnet
states~\cite{Zhang06,Abanin07-1,Jiang07,Abanin07-2} can also be
considered for the counter-propagating edge states. But, the Zeeman
energy in 10 T is only about 20 K,~\cite{Abanin07-2} which is much
smaller than the energy difference between the lowest and the first
exited Landau levels, $\sim$700 K.~\cite{Novoselov07} Although this
phenomenon can be understood by assuming both the incomplete mode
mixing and the presence of counter-propagating edge
states,~\cite{Chklovskii92,Rossier07,Silvestrov08,Zhang06,Abanin07-1,Jiang07,Abanin07-2}
further theoretical and experimental investigation is required for a
conclusive account for the effect.

\section{SUMMARY}

We studied the edge-state equilibration processes in a locally
gated \textit{p-n-p} junction of the graphene by measuring the
four-terminal QH resistances. Our scheme enables measurements of
finite longitudinal and asymmetric diagonal QH resistances, which
furnish precise information on the reflected as well as the
transmitted QH edge states. Most of our observations are
quantitatively analyzed by the Landauer-B\"{u}ttiker
formula~\cite{Beenakker91,Buttiker88} based on the complete-mode
mixing hypothesis. But, unexpected resistance plateaus
corresponding to the 5/2 transmission of edge states are also
observed. We suggest that it may arise from the incomplete mode
mixing and/or the presence of the counter-propagating QH edge
states.~\cite{Chklovskii92,Rossier07,Silvestrov08,Zhang06,Abanin07-1,Jiang07,Abanin07-2}

Our simple four-terminal measurement scheme reported here is not
sensitive to the contact resistance and the sample geometry, so
that it can be conveniently used for the edge-state manipulation
of arbitrary-shaped devices such as Mach-Zehnder
interferometers.~\cite{Ji03} Moreover, with the additional
information on the reflection of the edge states, this scheme
enables one to investigate the details of the edge-state
equilibration at \textit{p-n} interfaces such as 5/2 transmission
of the edge states found in this study. It can be also employed to
investigate the edge states in hybrid structures of mono-layer and
multi-layer graphene.~\cite{Nilsson07,Puls08}
\\

\begin{acknowledgments}
Critical reading of the paper by L. Paulius is deeply appreciated.
We are grateful for the valuable discussion with Y.-W. Son and
K.-S. Park. This work was supported by Acceleration Research Grant
No. R17-2008-007-01001-0 by Korea Science and Engineering
Foundation.
\end{acknowledgments}

\end{document}